\documentclass[twocolumn,prl,aps,superscriptaddress,floatfix,showpacs]{revtex4}
\usepackage{bm}
\usepackage{epsfig}

\begin{document}

\title{Spin dynamics of the ferromagnetic superconductor UGe$_{2}$}

\author{S. Raymond}
\affiliation{CEA-Grenoble, DSM / DRFMC / SPSMS, 38054 Grenoble, France}
\affiliation{Institut Laue Langevin, 38042 Grenoble, France}
\author{A. Huxley}
\affiliation{CEA-Grenoble, DSM / DRFMC / SPSMS, 38054 Grenoble, France}

\begin{abstract}
Inelastic neutron scattering was used to study the low energy magnetic excitations of the ferromagnetic superconductor UGe$_{2}$. The ferromagnetic fluctuations are of Ising nature with a non-conserved magnetization and have an intermediate behavior between localized and itinerant magnetism. 
\end{abstract}


\maketitle

Superconductivity occuring close to an antiferromagnetic quantum critical point is now well documented. In the magnetically mediated model for superconductivity, the formation of Cooper pairs is also expected close to a low temperature ferromagnetic transition. But it is only very recently that the first experimental evidence for the coexistence of ferromagnetism and bulk superconductivity was found in UGe$_{2}$ under pressure \cite{Saxena,Huxley1}. At 12 kbar, the superconducting transition temperature is maximum ($T_{c}  \approx$ 0.7 K) while the ferromagnetism remains strong with $T_{Curie} \approx $ 30 K and a magnetic moment of 1 $\mu_{B}$. Given the large internal field, triplet superconductivity is likely. In order to get a microscopic insight into this new physics, we have performed inelastic neutron scattering (INS) experiments on a single crystal of UGe$_{2}$ using the cold three axis spectrometer IN14 at the Institut Laue Langevin, France. UGe$_{2}$ crystallizes in the Cmmm space group with $a$=3.99 $\AA$, $b$=15.04 $\AA$ and $c$=4.09 $\AA$.

\begin{figure}[h]
\epsfig{file=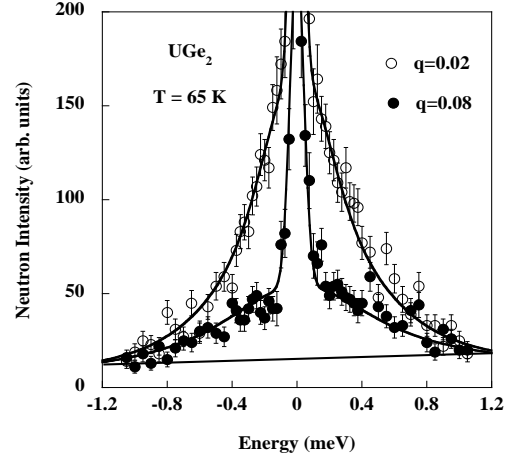,height=65mm,width=60mm,angle=90}
\caption{Energy response of the magnetic signal for $\bf{Q}$=(0,0,1.02) and  $\bf{Q}$=(0,0,1.08) at $T$=65 K. Lines correspond to fit to Lorentzian functions as described in the text. The lower  solid line denotes the background estimated from measurements performed at 1.5 K. The narrow central peak corresponds to incoherent background.}
\label{figure1}
\end{figure}

Quasielastic scattering corresponding to critical and paramagnetic scattering was measured in the range 20-70 K. Typical spectra obtained at $T$=65 K are shown in  Fig.\ref{figure1} for $\bf{Q}$=(0, 0, 1.02) and $\bf{Q}$=(0, 0, 1.08) (Here, $\bf{Q}$=$\bf{\tau}$+$(q_{h},q_{k},q_{l})$, where $\bf{\tau}$ is a Bragg peak and each coordinate expressed in reciprocal lattice units). Most data were taken along the c-axis and in the following, we write $q=q_{l}$. The data were fitted using the double Lorentzian form : 
\begin{equation}
   {\chi''(q,\omega)}= \frac{\chi_{0}}{1+( q/\kappa)^{2}}\frac{\omega\Gamma_{q}}{\omega^{2}+{\Gamma_{q}}^{2}}   
\label{equation}
\end{equation}
with $\chi_{0}$ the bulk susceptibility, $\kappa$ the inverse correlation length and $\Gamma_{q}$ the relaxation rate. The first fraction in (1) is the $q$-dependent susceptibility $\chi_{q}$. Details of the data analysis and results are given elsewhere \cite{Huxley2}.  For each temperature, $\Gamma_{q}$ is directly obtained  from constant $q$ scans ; $\kappa$  and $\chi_{0}$ are deduced by a fit of $\chi_{q}$ (obtained by integrating over energy $\chi''(q,\omega)/\omega$).   Here, we emphasize the main points \cite{Huxley2}.  i) Only longitudinal fluctuations  were observed confirming the  Ising nature of the system. ii) These fluctuations are observed both above and below $T_{Curie}$ and no new Goldstone modes are observed below $T_{Curie}$. iii) the product $\Gamma_{q}\chi_{q}$ is independent of temperature above $T_{Curie}$ but drops when $T$ decreases below $T_{Curie}$. iv) Except at $T_{Curie}$, it is found that the magnetization is not conserved, in other words the relaxation rate $\Gamma_{q}$ is finite in the limit $q \approx 0$. In this short paper, we focus on the relationship above $T_{Curie}$ between the relaxation rate and the inverse correlation length and compare this with the predictions for dynamical scaling theory for kinetic Ising model with a non-conserved magnetization \cite{Hohe}.
\begin{figure}[h]
\epsfig{file=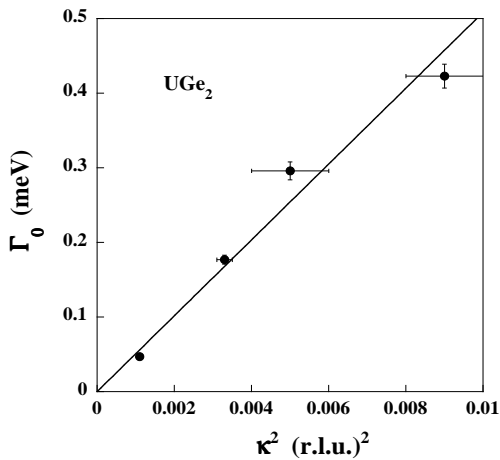,height=65mm,width=60mm,angle=90}
\caption{$\Gamma_{q}$ at $q$=0 as a function of $\kappa^{2}$, where temperature is an implicit parameter. The line is a linear fit to the data.}
\label{scaling0}
\end{figure}
In \cite{Huxley2}, it was shown that $\chi_{0}\kappa^{2}$ is independent of temperature. Above $T_{Curie}$, $\chi_{0}\Gamma_{0}$ is also temperature independent (at $q$=0 as for any $q$ value). We then have $\Gamma_{0}=a\kappa^{2}$. Such a relation is shown in Fig.\ref{scaling0} where $T$ is an implicit parameter. The best fit is obtained with a=50(1) meV(r.l.u.)$^{-2}$. Such a relation is also found for a non-conserved Ising model \cite{Hohe} and implies that the dynamical exponent $z$ equals two in UGe$_{2}$ ($\Gamma_{0}=a\kappa^{z}$). The $q \neq$ 0 generalization of such a law is the scaling form $\Gamma_{q}=Aq^{z}f(\kappa/q)$. The result of  such an analysis based on  the collapse onto a single curve  of data obtained at 55, 60, 65 and 70 K is shown in Fig.\ref{scalingq}  with $f(x)=1+x^2$, $z$=2 and  $A$=49(1) meV(r.l.u.)$^{-2}$. There is a very good agreement between the determination of $a$ and $A$ ($a$ is determined from a subset of data with $q$=0). Such plots have also been made for metallic Heisenberg ferromagnets, but with $z$=2.5 as expected from dynamical scaling theory for that case \cite{Hohe}. The same scaling function $f$ was applied to MnSi \cite{Ishi}, CoS$_{2}$\cite{Hira} and Ni$_{3}$Al\cite{Semadeni}. In these compounds it was argued that the form of  $f$ is related to the itinerant character of the system and contrasts with the form expected for localized systems where $f$ is experimentally found to be close to that calculated by  R\'esibois-Piette \cite{Resi}. 
For UGe$_{2}$, the $z$ and $f$ used to describe the data simply correspond to $\Gamma_{q}=A(\kappa^{2}+q^{2})$, with all the temperature dependence included in $\kappa$. 
\begin{figure}[h]
\epsfig{file=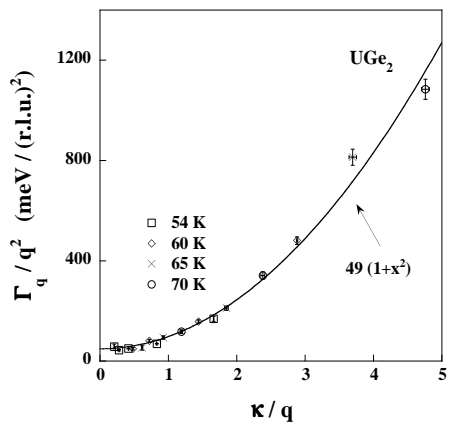,height=55mm,width=60mm,angle=0}
\caption{$\Gamma_{q}/q^{2}$ as a function of $\kappa/q$. The line is a fit to the data using dynamical scaling theory as explained in the text.}
\label{scalingq}
\end{figure}

In the study of the Heisenberg ferromagnets, focus was also given to the quantity $\kappa_{0}$ defined by the temperature variation of the inverse correlation length, $\kappa(T)^{2}=\kappa_{0}^{2}(T/T_{Curie}-1)^{2\nu}$ above $T_{Curie}$. For UGe$_{2}$, we found $\nu$=0.5 and $\kappa_{0}^{2} \approx$ 0.08  $\AA^{-2}$ \cite{Huxley2}. The quantities $\kappa_{0}$ and $A$ characterize the extension of the spin fluctuations in  $q$ and $\omega$ space respectively. In itinerant systems, the extension is narrow in $q$ space and wide in $\omega$ space \cite{Mori}. To this respect, $\kappa_{0}^{2}$ is small as in the itinerant system MnSi ( 0.035 $\AA^{-2}$ \cite{Ishi}) while it is generally larger in localized systems (0.41 $\AA^{-2}$ in EuO, 1.1 $\AA^{-2}$ in Fe). The ratio $A/T_{Curie}$ is also usually used to describe the energy spread of spin fluctuations \cite{Hira,Mori}.  For UGe$_{2}$, we found $A/T_{Curie} \approx$ 0.9 meV/K. In fact to compare with Heisenberg systems, $A$ must be re-scaled to $A'$ in order to take into account the difference in $z$ between kinetic Ising and Heisenberg systems (e.g. $A'q_{ZB}^{2.5}=Aq_{ZB}^{2}$ where $q_{ZB}$ is the zone boundary wave-vector). After such a phenomenological correction, the ratio  $A'/T_{Curie}$ for UGe$_{2}$ (1.3 meV/K) is about twice the one of EuO (0.77 meV/K) and half the one of MnSi (3.3 meV/K). This intermediate behavior between localized and itinerant magnetism was also pointed out in the study of the temperature variation of the order parameter for several pressures \cite{Huxley3}.

\vspace{5mm}

The spin dynamics of UGe$_{2}$ is of fundamental interest for magnetism by itself since there are a few detailed INS studies on Ising ferromagnets \cite{Collins}. Considering UGe$_{2}$ as such a model system above $T_{Curie}$, the prediction of dynamical critical phenomena for the kinetic Ising model with a non-conserved magnetization are experimentally verified as shown by the scaling of the neutron linewidth implying a dynamical exponent, $z$=2. As concerns superconductivity,  microscopic parameters like $\Gamma_{q}$ and $\kappa_{0}$ are necessary inputs in the theory of magnetically mediated superconductivity \cite{Monthoux}. The Ising fluctuations favor triplet superconductivity together with the rather low value of $\kappa_{0}$. However the evolution of these parameters as a function of pressure in the range where superconductivity occurs has to be measured. 

\vspace{5mm}
We acknowledge ILL staff for technical support and discussions concerning optimal use of instrument with A. Stunault.

\end{document}